\documentclass[smallextended]{svjour3}       
\usepackage[utf8]{inputenc}
\usepackage{amsfonts}
\usepackage{amsmath}
\usepackage{amssymb}
\usepackage[pdftex]{graphicx}

\usepackage{xcolor}

\smartqed  

\title{A lattice QCD study of pion-nucleon scattering in the Roper channel}

\titlerunning{Pion-nucleon scattering in the Roper channel}        

\author{Luka Leskovec \and Christian B. Lang \and M. Padmanath \and Sasa Prelovsek}

\authorrunning{Luka Leskovec} 

\institute{Luka Leskovec \at
            Department of Physics, University of Arizona,\\
            Tucson, AZ 85721, USA\\
            \email{leskovec@email.arizona.edu}
            \and
            Christian B. Lang \at
		    Institute of Physics, University of Graz,\\
            A-8010 Graz, Austria
            \and
            M. Padmanath \at
            Instit\"ut  f\"ur  Theoretische Physik, Universit\"at Regensburg,\\
            D-93040 Regensburg, Germany
            \and
            Sasa Prelovsek \at
            Instit\"ut  f\"ur  Theoretische Physik, Universit\"at Regensburg,\\
            D-93040 Regensburg, Germany\\
            Faculty of Mathematics and Physics, University of Ljubljana,\\
            1000 Ljubljana, Slovenia\\
            Jozef Stefan Institute,\\
            1000 Ljubljana, Slovenia
}

\date{Received: date / Accepted: date}

\begin{document}
\maketitle
\begin{abstract}
We present a lattice QCD study of the puzzling positive-parity nucleon channel, where the Roper resonance
$N^*(1440)$ resides in experiment. The study is based on an ensemble of gauge configurations
with $N_f=2+1$ Wilson-clover fermions with a pion mass of $156$ MeV and lattice size
$L=2.9$ fm. We use several $qqq$ interpolating fields combined with $N\pi$ and $N\sigma$ two-hadron operators
in calculating the energy spectrum in the rest frame. Combining experimental $N\pi$ phase shifts with elastic
approximation and the L\"uscher formalism suggests  in the spectrum an additional energy level near the Roper
mass $m_R=1.43$ GeV for our lattice. We do not observe any such additional energy level, which implies that
$N\pi$ elastic scattering alone does not render a low-lying Roper resonance. The current status indicates
that the $N^*(1440)$ might arise as dynamically generated resonance from coupling to other channels, most
notably the $N\pi\pi$.
\end{abstract}

\section{Motivation}
The baryonic sector of hadrons, composite particles made up from quarks and gluons, holds many
interesting states. One of the most prominent of them is the Roper resonance $N^*(1440)$
with quantum numbers $(I)J^P=(1/2)1/2^+$, like the nucleon. The $N^*(1440)$ is a strongly unstable
baryon with a decay width $\Gamma \approx 300$ MeV. Experimentally it has been observed to couple to
the $N\pi$ and $N\pi\pi$ channels. In the latter, however, several resonances such as the $\rho$,
$\Delta$ and the exotic $\sigma$ can appear; phenomenologically these couplings are associated to
the meson cloud contribution to the excited nucleon, $N^*(1440)$.

The riddle of the Roper resonance, i.e., why is it so light, arises when we
compare  the $J=1/2$ baryon spectrum with that of an hydrogen atom. The masses
of the hadrons then depend on the radial and orbital quantum numbers $n_r$ and
$l$ . The radial excitation relates to the number of radial nodes in the wave
function and the latter is the angular momentum quantum number.  In this picture
the energy levels are $E_{n_r,l} = \Omega_0 ( (n_r + l) + 3/2)$ ; the ground
state  $(n_r=0,l=0)$ is the nucleon with $J^P=1/2^+$ followed by the second
state, $(n_r=1,l=1)$, called $N^*(1535)$ with $J^P=1/2^-$. The third state then
appears in the $J^P=1/2^+$ channel above the $N^*(1535)$, supposedly the Roper
resonance. However experimental measurements find an unconventional level
ordering with the Roper mass below the $N^*(1535)$, a phenomenon not yet
completely understood \cite{Burkert:2017djo}.

Previous lattice studies have attempted to determine the Roper mass using several different
approaches, however sharing a common feature. They all used an approach, where the
spectrum was determined with $qqq$ interpolating fields  \cite{Alexandrou:2013fsu,Alexandrou:2014mka,Engel:2013ig,Edwards:2011jj,Mahbub:2013ala,Roberts:2013ipa,Liu:2016rwa,Wu:2017qve}. The only lattice calculation that
included five quark interpolators used strictly local $qqqq\bar{q}$ interpolators \cite{Kiratidis:2016hda},
which seem to couple weakly with multi-hadron states in practice.
In our study we applied a different approach, where we used (local) single hadron as well as
(non-local) two hadron interpolating operators in calculating the spectrum of the
$J^P=1/2^+$ channel.

\section{Hadron Spectroscopy with lattice QCD}

A lattice QCD calculation of the spectrum is performed in a finite volume box with a spatial size $L$ and
periodic boundary conditions in space.
 This has several implications: (i) The continuum
 symmetry group $O$ is reduced to the double-covered orthogonal cubic group and the
 continuum rotations are restricted to the corresponding irreducible representations.
(ii) Due to
the periodic boundary conditions, the energy spectrum becomes discrete. Because of that the
completeness relation is modified from
\begin{align}
I = \sum_{n \in D} |n\rangle \langle n| + \int_{CS} d\alpha |\alpha\rangle\langle\alpha|
\end{align}
in the infinite volume to
\begin{align}
\label{CR:box}
I = \sum_{n \in D} |n\rangle \langle n| + \sum_{DS} |m\rangle\langle m|
\end{align}
in the finite volume. Above, $D$ denotes the discrete states,$CS$ denotes the
continuum scattering states and $DS$ denotes the discrete scattering states. Because of this we
cannot separate discrete states from the scattering states in the finite volume. Most importantly
the spectrum determined with lattice QCD will contain all states (those associated with scattering
states as well as resonances) with the proper quantum numbers.

To determine the spectrum we use several interpolating operators: (i) (local) single hadron operators;
these are interpolating fields made up from three quarks ($qqq)$ and jointly projected to a definite momentum;
and (ii) (non-local) two hadron operators, where the separate hadrons, either a baryon ($qqq)$ or a meson ($\bar qq$), are
projected to momentum separately. From these interpolating operators we then build a $2$-point
correlation matrix:
\begin{align}
C_{ij}(t) = \langle 0|O_i(t)O^{\dagger}_j(0)|0\rangle.
\end{align}
By inserting the completeness relation from Eq. \ref{CR:box} and propagation, $\exp[-H t]$, to time $t$
the correlation matrix is decomposed as:
\begin{align}
C_{ij}(t) = \sum_{n \in D} Z_i^n Z_j^{n\*} e^{- E_n t} + \sum_{m \in DS} Z_i^m Z_j^{m\*} e^{- E_m t}.
\end{align}
We can easily see that the discrete states (bound states and resonances) and the scattering states cannot be disentangled
in general. However, based on the various types of interpolating operators used, information about their
likely nature can be inferred from the overlap factors $Z_i^n$.

\begin{figure}[htb!]
\centering
\includegraphics[width=0.35\textwidth]{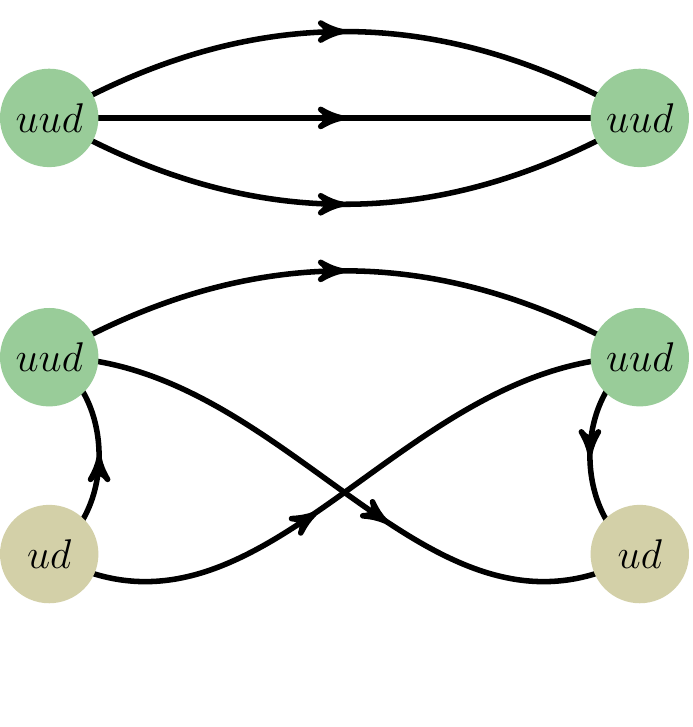}
\caption{\label{fig:wick} Examples of Wick contractions related to two-hadron spectrum in the
$J^P=1/2^+$ channel. The green circles represent the nucleon and grey circles the mesons $\pi$ and $\sigma$. The black 
lines connecting the sink (left side) and the source (right side) are the light quark propagators.}
\end{figure}

We build the correlation matrix $C_{ij}(t)$ using quark propagators to build all the needed Wick
contractions. In our case, that means Wick contractions connecting $3$- and $5$-quarks sources with $3$- or $5$-quark sinks.
All in all there are $84$ Wick contractions involved in building our correlation matrix
and some examples are shown in Fig.~\ref{fig:wick} and the rest can be found in Ref. \cite{Lang:2012db}.

The energy levels of the spectrum are determined by fitting the principal correlators $\lambda_n(t,t_0)$
determined with the variational approach:
\begin{align}
C_{ij}(t)u_j^n = \lambda_n(t,t_0) C_{ij}(t_0)u_j^n,
\end{align}
where in the limit $\lim_{t\to\infty}\lambda_n(t,t_0) \propto e^{-E_n t}$.

\begin{figure}[htb!]
\centering
\includegraphics[width=1.0\textwidth]{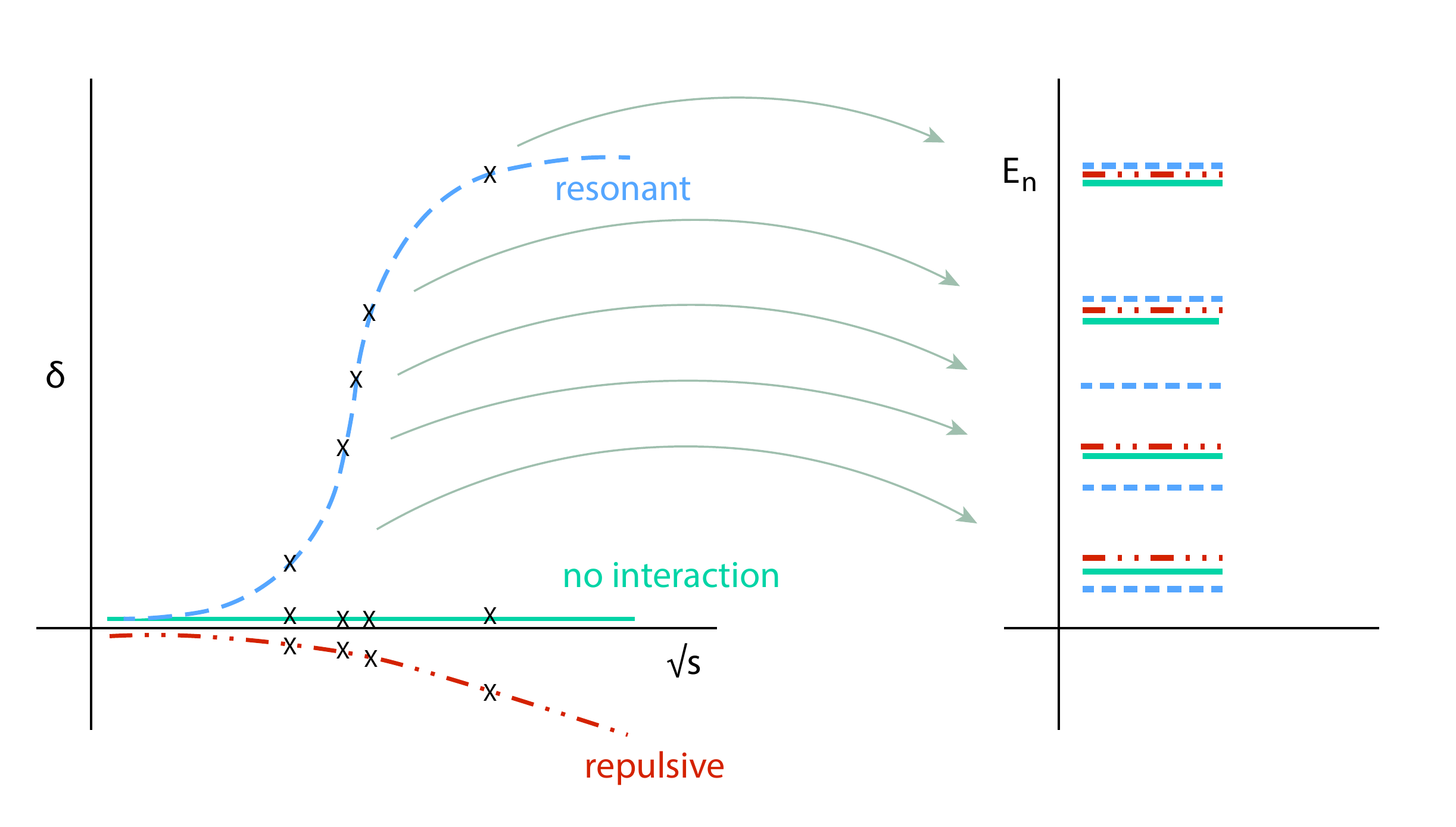}
\caption{\label{fig:luscher} Schematic representation of the three possible phase shift scenarios in elastic scattering and their corresponding
finite volume spectra as determined by the L\"uscher method. The full green line represents the case where there is no 
interaction between the two hadrons, red dot-dot-dashed line represents a repulsive interaction between the two hadrons 
and the dashed blue line a resonant interaction.}
\end{figure}

When considering two hadron systems on the lattice the discrete spectrum in the finite volume is
analytically related to the infinite volume elastic phase shift $\delta$ via a mapping
first derived by L\"uscher \cite{Luscher:1990ux} and recently reviewed in Ref. \cite{Briceno:2017max}.
However, this approach can also be inverted: if we know the elastic phase shift  we can determine the expected
spectrum in a finite volume. We consider three cases for elastic scattering here:  (i) no interaction between the baryon and
meson, (ii) repulsive scattering of a baryon and meson  and (iii) resonant scattering of a baryon and
a meson. A schematic representation of the three different situations is shown in Fig.~\ref{fig:luscher}.

In the case of no interaction (case i), shown as the full green line in Fig.~\ref{fig:luscher}, we find the spectrum to be made up of scattering levels, whose
energies correspond to
\begin{align}
\label{eq:noint}
E_{no-int} = \sqrt{m_{N}^2 + \vec{p}_N^2} + \sqrt{m_{\pi}^2 + \vec{p}_\pi^2},
\end{align}
where $m_N$ is the nucleon mass, $m_{\pi}$ is the pion mass and
$\vec{p}_i = \frac{2\pi}{L}\vec{n}$, $\vec{n}\in\mathbb{Z}^3$ for $i=N,\pi$. The presence of any kind
of interaction, repulsive or attractive, between the two hadrons leads to an energy shift with respect
to the non-interacting energies in Eq. \ref{eq:noint}. When the interaction is repulsive (case ii), as shown by the dot-dot-dashed red line in Fig.~\ref{fig:luscher},
the energies shift slightly with respect to the non-interacting ones, typically to just slightly higher
values. However, when the interaction is attractive enough to produce a resonance such as in (case iii)
the spectrum changes significantly in comparison to the non-interacting case. The main change is the
appearance of an additional energy level near the expected resonance mass; additionally we also find
the other energy levels have moved in a direction away from the additional
level. That is energies below the additional energy level shift toward lesser values while the energy levels
above shift to greater values.
Thus, a general finding when the elastic hadron scattering is resonant is the appearance of an
additional energy level near the resonance mass. An illustration of a resonant phase shift and the additional energy level is shown by the dashed blue lines in Fig.~\ref{fig:luscher}.

\section{Lattice parameters}
Our calculations were performed on a lattice with $L=2.90$ fm and light quark masses corresponding
to $m_{\pi}=156(7)$ MeV and $m_N=955(12)$ MeV. The $N_f=2+1$ dynamical quarks as well as the valence
quarks are implemented as clover improved wilson fermions \cite{Aoki:2008sm}. To evaluate the correlators we used
the full distillation approach \cite{Peardon:2009gh} which allows for a computationally efficient way to calculate
the many partially-disconnected diagrams that appear in this channel.

\section{The Roper as a vanilla resonance}
The Roper resonance appears in the $J^P=1/2^+$ $N\pi$ scattering channel with a resonance mass
$m_{N^*(1440)}\approx 1430$ MeV and a decay width of $\Gamma \approx 350$ MeV. Experimentally its
phase shift starts to rise at approximately $1.2$ GeV and reaches $180^\circ$ around $1.7$ GeV.
The inelasticity parameter that measures the coupling to other channels, e.g. $N\pi\pi$,
is far from constant and exhibits a large change in the region around $1.4$ GeV.

Ideally one should take into account non-elasticity in a multi-channel analysis. This
is not possible (yet) since for this we would need more lattice volumes and data \cite{Briceno:2017max}.
We therefore
work with  thehypothesis that the $N^*(1440)$ arises as a resonance in
elastic $N\pi$ scattering. We also assume that our basis to construct the correlation matrix
$C_{ij}$ is sufficiently large and diverse to have a significant overlap to all relevant states
and that a chiral implementation of fermions on the lattice is not crucial to obtain a sufficient
overlap.

\begin{figure}[htb!]
\centering
\includegraphics[width=0.9\textwidth]{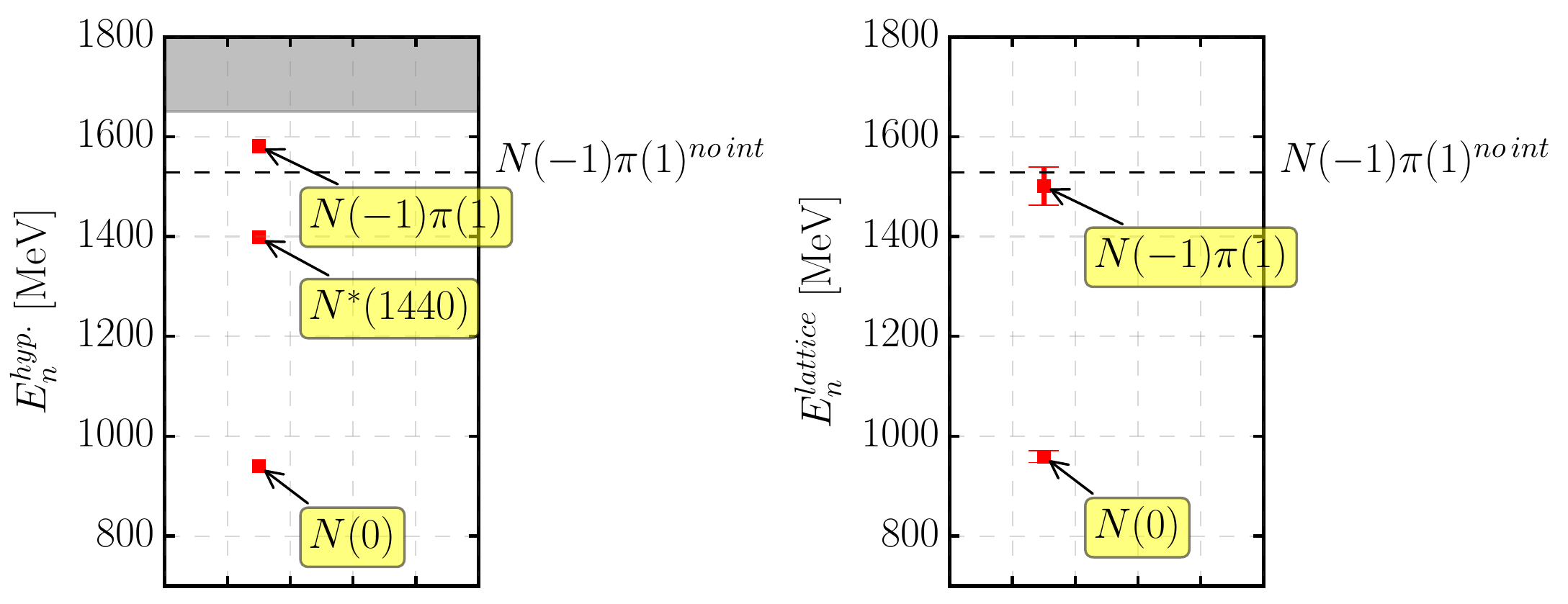}
\caption{\label{fig:spec} The dashed lines indicate the non-interacting energies and the red squares the finite volume energies. 
Comparison between the spectrum in the $J^P=1/2^+$ channel obtained from experimental phase shifts (left panel) and 
the spectrum obtained from a lattice QCD calculation based on $qqq$ and $N\pi$ operators (right panel). The discrepancy 
between the spectra clearly demonstrates that the Roper is not a vanilla resonance in elastic $N\pi$ scattering.}
\end{figure}

We compare the spectrum calculated with lattice QCD with the expected spectrum obtained from the
experimental phase shift in Fig.~\ref{fig:spec}.   If the Roper was a vanilla resonance\footnote{Like
for example the $\rho$ meson.} in $N\pi$ scattering, then we would observe an additional energy
level as shown in the left panel of Fig.~\ref{fig:spec}. If it was however a result of dynamical coupling
between the $N\pi$ and $N\pi\pi$ channel (or possibly even $N\eta$), or if any other of our
assumptions were not correct, then the additional energy level might not
appear. The right panel of  Fig.~\ref{fig:spec} shows our  lattice spectrum obtained from the
correlation matrix based on $qqq$ and $N\pi$ interpolating operators. The additional level is absent
leading to the finding that the Roper cannot be a vanilla resonance resulting from only
$N\pi$ scattering.

This however does not mean, the Roper does not exist, but rather that the experimental Roper state might arise from
dynamical coupling to a three particle channel; while this is an active field of research
\cite{Briceno:2017tce,Hammer:2017kms,Mai:2017bge}, the framework to study the Roper has not yet been
developed.

\section{Discussing the spectrum}
\begin{figure}[htb!]
\centering
\includegraphics[width=0.9\textwidth]{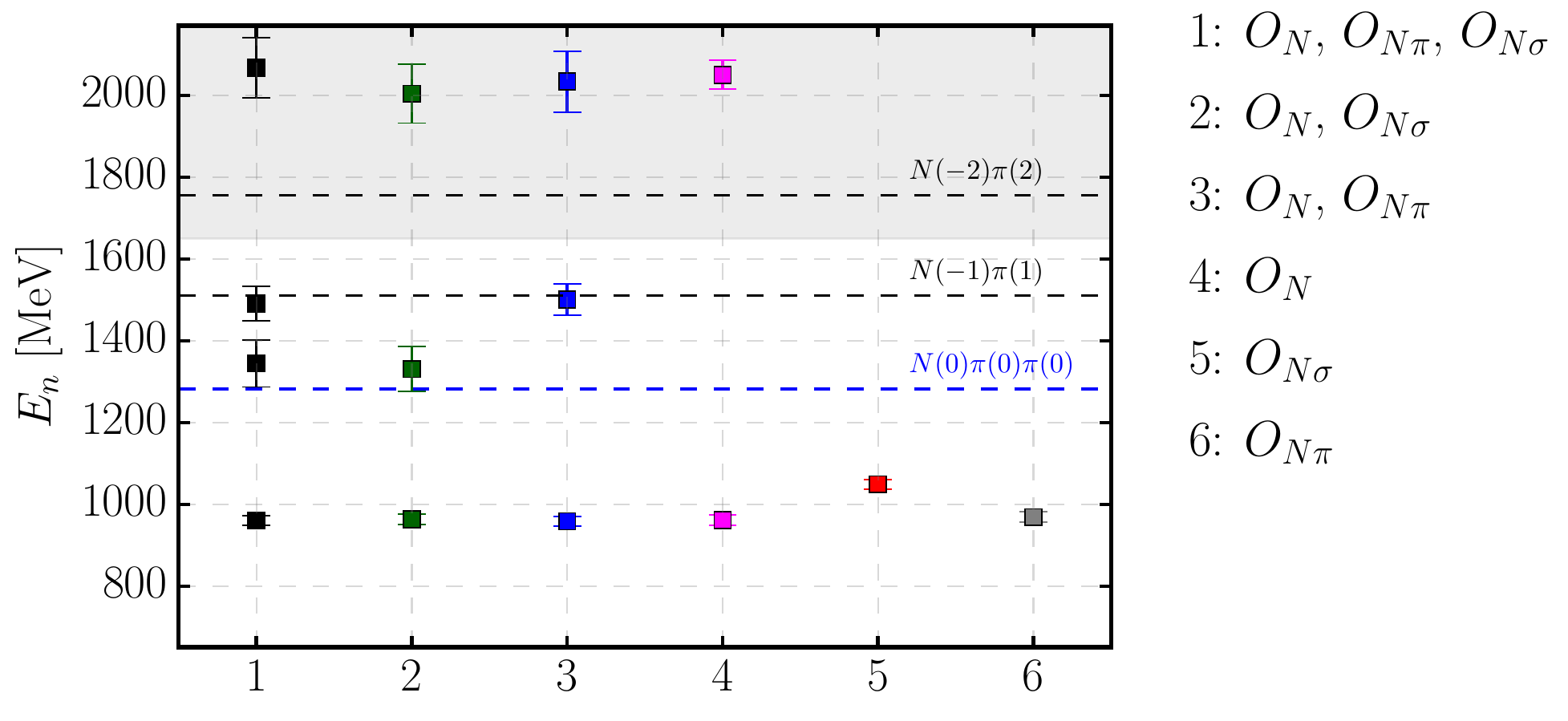}
\caption{\label{fig:allspec} Spectra for various bases considered. The black dashed lines represent the non-interacting $N\pi$ energy 
levels while the dashed blue line indicates the energy of the $N\pi\pi$ threshold.}
\end{figure}

To better understand this channel we continue by adding also $N\sigma$ interpolating
fields to the correlation matrix; the $\sigma$ couples to $\pi\pi$ in $s$-wave thus gives us (limited)
access to the $N\pi\pi$ channel. The results for the various different bases are shown in Fig.~\ref{fig:allspec}
Our interpolating operators were designed to cover the energy region up to $1.65$ GeV and thus any
energy levels lying above that are not fully reliable, i.e. can be related to an unknown mixture of states. When we
consider only $N$ or $N\pi$ interpolating fields we find the ground state energy to be consistent with the
nucleon mass and when considering only the $N\sigma$ interpolating fields, the ground state is somewhat
higher than the  nucleon mass, which is possibly explained by the interpolator having a bad coupling
to the ground state and being
a linear combination of several states. When nucleon and $N\pi$ interpolators are considered we
find results already discussed in the previous section; however when we replace the $N\pi$
interpolator with the $N\sigma$ one we find the first excited state has moved to a lower energy;
one consistent with the $N\pi\pi$ threshold energy. When all, nucleon,
$N\pi$ and $N\sigma$, interpolators are included we find that the spectrum contains energy levels
consistent with the nucleon mass, the $N\pi\pi$ threshold and lowest $N\pi$ non-interacting scattering energy
below $1.65$ GeV. No additional energy levels are present. However, as we do not yet know what kind of spectrum we
expect if the Roper was generated dynamically via coupled channel scattering, we cannot conclude
anything about the Roper based on the spectrum alone.
For lattice QCD to provide any input on this puzzle, a more complicated and involved analysis is needed.\\

Comparing our spectrum results with model studies confirms our findings. In particular when comparing
to Ref. \cite{Liu:2016uzk}, our lattice energy levels below  $1.65$ GeV disagree with a only bare
Roper $qqq$ core interpretation, but are consistent with results when the $N^*(1440)$ resonance
is generated dynamically from coupling between the $N\pi$, $N\sigma$ and $\Delta\pi$ channels.
Our findings also agree with a recent model study in Ref. \cite{Golli:2017nid}, where the Roper arises as
a pole in the scattering matrix via dynamical coupling to the $N\pi$ and $N\pi\pi$ channels.

\section{Conclusions}
We performed a lattice QCD calculation of the $J^P=1/2^+$ channel using (local) single hadron and (non-local)
two hadron interpolating fields on a ensemble of gauge fields with $m_{\pi}\approx 156$ MeV \cite{Lang:2016hnn}.
We found
that the Roper cannot arise as a resonance in elastic $N\pi$ scattering and is likely a consequence
of dynamical coupling between several channels. We also note, that the spectrum might be different if we had used
lattice fermions with a better chirality or pentaquark-like
interpolating fields. Further spectrum and structure studies of this channel using lattice QCD are required to
understand this resonance.

\section{Acknowledgments}

We thank the PACS-CS collaboration for providing the gauge configurations. We also thank M. D\"oring, L. Glozman, Keh-Fei Liu, D. Mohler, B. Golli, M. Rosina and S. Sirca for valuable discussions. We are grateful to  for numerous valuable discussions and suggestions. This work is supported in part by the Slovenian Research Agency ARRS, by the Austrian Science Fund FWF:I1313-N27 and by the Deutsche Forschungsgemeinschaft Grant No. SFB/TRR 55. M. P. acknowledges support from EU under grant no. MSCA-IF-EF-ST-744659 (XQCDBaryons). The calculations were performed on computing clusters at the University of Graz (NAWI Graz) and Theoretical Department at Jozef Stefan Institute. 


\providecommand{\href}[2]{#2}\begingroup\raggedright\endgroup

\end{document}